\documentclass[conference]{IEEEtran}
\pdfoutput=1
\usepackage{amsfonts}
\usepackage{graphicx}
\usepackage[cmex10]{amsmath}
\usepackage{amsthm}
\IEEEoverridecommandlockouts
\newtheoremstyle{dotless}{}{}{}{ }{\it}{:}{ }{}

\theoremstyle{dotless}
\usepackage{amssymb,amsthm,amsmath,amsfonts}
\newtheorem{proposition}{Proposition}

\newtheorem{lemma}{Lemma}
\newcommand{\beq}{\begin{equation}}
\newcommand{\eeq}{\end{equation}}

\begin{document}
%
\title{Transmit Beamforming for MIMO Communication Systems with Low Precision ADC at the Receiver}

\author{\IEEEauthorblockN{Tapan Shah}
\IEEEauthorblockA{School of Technology and Computer Science\\
Tata Institute of Fundamental Research\\
Mumbai, India\\
Email: tapanshah@tifr.res.in}
\and
\IEEEauthorblockN{Onkar Dabeer}
\IEEEauthorblockA{School of Technology and Computer Science\\
Tata Institute of Fundamental Research\\
Mumbai, India\\
Email: onkardabeer@gmail.com}}


%


\maketitle

\begin{abstract}
Multiple antenna systems have been extensively used by standards designing multi-gigabit communication systems operating in bandwidth of several GHz.  In this paper, we study the use of transmitter (Tx) beamforming techniques to improve the performance of a MIMO system with a low precision  ADC. We motivate an  approach to use eigenmode transmit beamforming (which imposes a diagonal structure in the complete MIMO system) and use an eigenmode power allocation which minimizes the uncoded BER of the finite precision system. Although we cannot guarantee optimality of this  approach, we observe that even low with precision ADC, it performs  comparably to  full precision system with no 
eigenmode power allocation. For example, in a  high throughput MIMO system with a finite precision ADC at the receiver, simulation results show that for a $\frac{3}{4}$ 
LDPC coded  $2\times 2$ MIMO OFDM 16-QAM system with 3-bit precision ADC at the receiver, a BER of $10^{-4}$ is achieved at an SNR of $26$ dB. This is $1$ dB better than that required for the same system with full precision but equal eigenmode power allocation.
\end{abstract}

\IEEEpeerreviewmaketitle

\section{Introduction}
Several standards designing muliGigabit communication systems (for example, IEEE 802.11ad and IEEE 802.15.3c) use multiple antennas at both the transmitter and receiver to boost up the data rates in the range of several Gbps. This gives rise to multiple input multiple output (MIMO) channel configurations. Almost all communication system with MIMO channels implement their receiver operations  in the digital domain and thus analog to digital converters (ADC) becomes a critical component for such systems. Most communication system use ADCs with a precision of 6--8 bits per sample. However,  high precision ADCs operating at sampling rates of several giga-samples-per second are extremely power hungry and expensive (\cite{ADCsurvey2008,walden,burmann}). Consequently, for designing  communication systems requiring such high speed sampling, ADC becomes a bottleneck. We would like to highlight that a similar problem does exist for digital-to-analog (DAC) conversion (DAC) at the 
transmitter. However, we assume that the transmitter has significantly more power resources compared to the receiver and we focus only on the ADC problem. An example of such a scenario is when a handheld device  downloads high definition content from an access point but uploads at normal speeds.

A naive method to reduce the power consumption at the receiver is to use an ADC with a low bit precision (1-4 bits per sample). However, this can lead to serious performance degradation (see Fig. \ref{fig:performance}). In the remainder of this section, we survey some of the previous works and highlight our contribution to improve the performance of a MIMO communication system when a low bit precision ADC is used at the receiver. We also set down some notational conventions at the end of this section.

\subsection{Prior Work}

Recently, there has been significant effort to address the ADC bottleneck and implement receivers  for multi-Gbps  single input single output (SISO) communication systems using low precision ADC at the receivers. Typically in  SISO OFDM systems, equal transmit power subcarrier (ETSP) power is used.  However, due to wide channel gain variations across the subcarriers, use of low precision ADC at the receiver results in loss of information from the weak carriers. As a result, the inter-carrier interference is not be canceled and we get an error floor (see Fig. 1 in \cite{tapan_vtc}). In \cite{tapan_vtc}, we suggest a transmitter based technique for subcarrier interference management using subcarrier power allocation to ameliorate the error floor. In \cite{tapan_tcom}, we further extend this work to find an optimal power allocation to minimize the error at the receiver. Using this optimal scheme, we observe that using a $3$--bit precision ADC at the receiver of $\frac{7}{8}$ LDPC coded $16$--QAM OFDM system, 
we can achieve a $2$ dB improvement in the performance compared to ETSP based OFDM. In this paper, we extend our earlier work to MIMO OFDM system.

There is considerable literature in the design of   transmitter-receiver (Tx-Rx) beamforming (joint or otherwise)\footnote{Classical beamforming often refers to a single beamvector at the transmitter. However, we consider a more generalized beamforming with multiple beamvectors. Some authors prefer to use the terms precoder and equalizer instead of Tx-Rx beamformers. For the sake of consistency, we will use the beamforming terminology.} techniques which optimizes a certain performance metric like mean square error (MSE), signal-to-interference noise ration (SINR), bit error rate (BER), transmit power etc. (see for example \cite{scaglione2002optimal, ding2003minimum,palomar2003joint, wang2000wireless} and references therein) for full precision MIMO receivers.   However, to the best of our knowledge, there has been no work in the design of Tx-Rx beamformers  for MIMO receivers for a finite precision ADC. 

In this paper, using ideas similar to the use of sub-carrier power allocation for OFDM systems to improve the performance of a low precision receiver (\cite{tapan_vtc}, \cite{tapan_tcom}), we use Tx-beamforming methods to  achieve full precision performance for a MIMO receiver with a low precision ADC.
 
 \subsection{Our Contribution}
 
Exact expressions for BER for finite precision  MIMO systems are fairly complicated and not amenable to finding closed form expression or computationally efficient algorithms for optimal Tx-beamformers. Instead, we  impose a specific structure on the Tx-beamformer which transmits on the eigenmodes and  diagonalizes the overall system. Although the {\em optimality of  diagonalization} property cannot be proved in general for a BER minimization criteria, we motivate this property from the existence of similar property for MSE minimization criteria. This greatly simplifies the optimization problem  and reduces it to a eigenmode power allocation problem. 

For such a diagonal structure, we compute  exact expression for the uncoded BER of the MIMO-OFDM system with finite precision ADC at the receiver (Proposition 1, part 1). Using this expression of uncoded BER, we obtain a eigenmode power allocation (OEPA) which minimizes it (Proposition 1, part 2). We also propose a useful closed form approximately optimal eigenmode power allocation \eqref{eq:approximation_infinity} which   can be easily used in practical system without significant increase in computational or storage requirements.  We use simulations to illustrate the improvement in the performance using our power allocation stream with a low precision ADC at the receiver.    As suggested in \cite{wpan}, we use the Saleh Valenzuela (SV)  to model the channel. We find that for a $\frac{3}{4}$ LDPC coded  $2\times 2$ MIMO OFDM 16-QAM system with 3-bit precision at the receiver, our method requires $1$ dB less power compared to the traditional full precision system with equal eigenmode power allocation (EEPA) 
to achieve a BER of $10^{-4}$. On the other hand, a 3-bit  system with EEPA has an error floor of $10^{-2}$.
 
 \subsection{Notation}
 
 We use small case bold face letter to represent vectors and small case italics letters to represent scalars. Upper case bold face letter are used to represent matrices. The superscripts $[\cdot]^\dagger$ and $[\cdot]^T$ are used to denote conjugate transpose and transpose, respectively. We use $\mathbf{X}=\text{diag}\left(\mathbf{X}_1\ldots\mathbf{X}_L\right)$ to represent a block diagonal matrix where each block is $\mathbf{X}_i$, $1\leq i\leq L$. We use $\mathbf{I}$ to denote an identity matrix. The dimension of the identity matrix follows from the context.
 
 \section{System Description}
 
 \subsection{MIMO Channel model}
 For each single input single output channel between transmit antenna $i$ and receive antenna $j$, we consider an independent ISI channel in which the resolved multipath components are grouped into $L_c$ clusters, each having $L_b$ rays. The time domain channel impulse response is given by
\begin{equation}
 h(t)=\sum_{c=0}^{H_{c}-1}\sum_{b=0}^{H_b-1}g_{c,b}\delta\left(t -T_{c}-\tau_{c,b}\right), \label{channel}
\end{equation}
where $g_{c,b}$ is the tap weight of the $b$-th ray of the $c$-th cluster, $T_c$ is the delay of $c$-th cluster, $\tau_{c,b}$ is the delay of the $b$-th ray relative to $T_c$ and $\delta(\cdot)$ is the dirac delta function. For simplicity of notation, we do not show the dependence on $i$ and $j$. Most standards like IEEE 802.15.3c which design communication system over a wideband channel, the Saleh-Valenzuela (S-V) model is the most popular model which characterizes the statistical properties of the parameters in \eqref{channel}. According to this model, 
\begin{align}
f\left(T_c|T_{c-1}\right)&=\Lambda\exp\left[-\Lambda\left(T_c-T_{c-1}\right)\right],\quad c>0\\
f\left(\tau_{c,b}|\tau_{c,\left(b-1\right)}\right)&=\lambda\exp\left[-\lambda\left(\tau_{c,b}-\tau_{c,\left(b-1\right)}\right)\right],\quad b>0
\end{align}
where $\Lambda$ and $\lambda$ are the cluster and ray arrival rate, respectively, and $f(\cdot)$ is the probability density function. Also, the mean square power of the tap weights are
\begin{equation}
\text{E}[|g_{c,b}|^2]=\text{E}[|g_{0,0}|^2]\exp\left(-\frac{T_c}{\Gamma}\right)\exp\left(-\frac{\tau_{c,b}}{\gamma}\right),
\end{equation}where $\Gamma$ and $\gamma$ are the cluster and ray decay rates, respectively. Since we are in the wideband regime, the distribution of the channel taps  is modeled by a lognormal  distribution \cite{ch_par,ch_par1}. 

The receiver implements a front end filter of sufficient bandwidth and then samples the received analog signal uniformly.  We assume that all the channel response vectors have length $L$

\subsection{Signal model for a  MIMO-OFDM channel with a finite precision receiver}

\begin{figure}[!ht]
\centering 
 \includegraphics[width=3.8in]{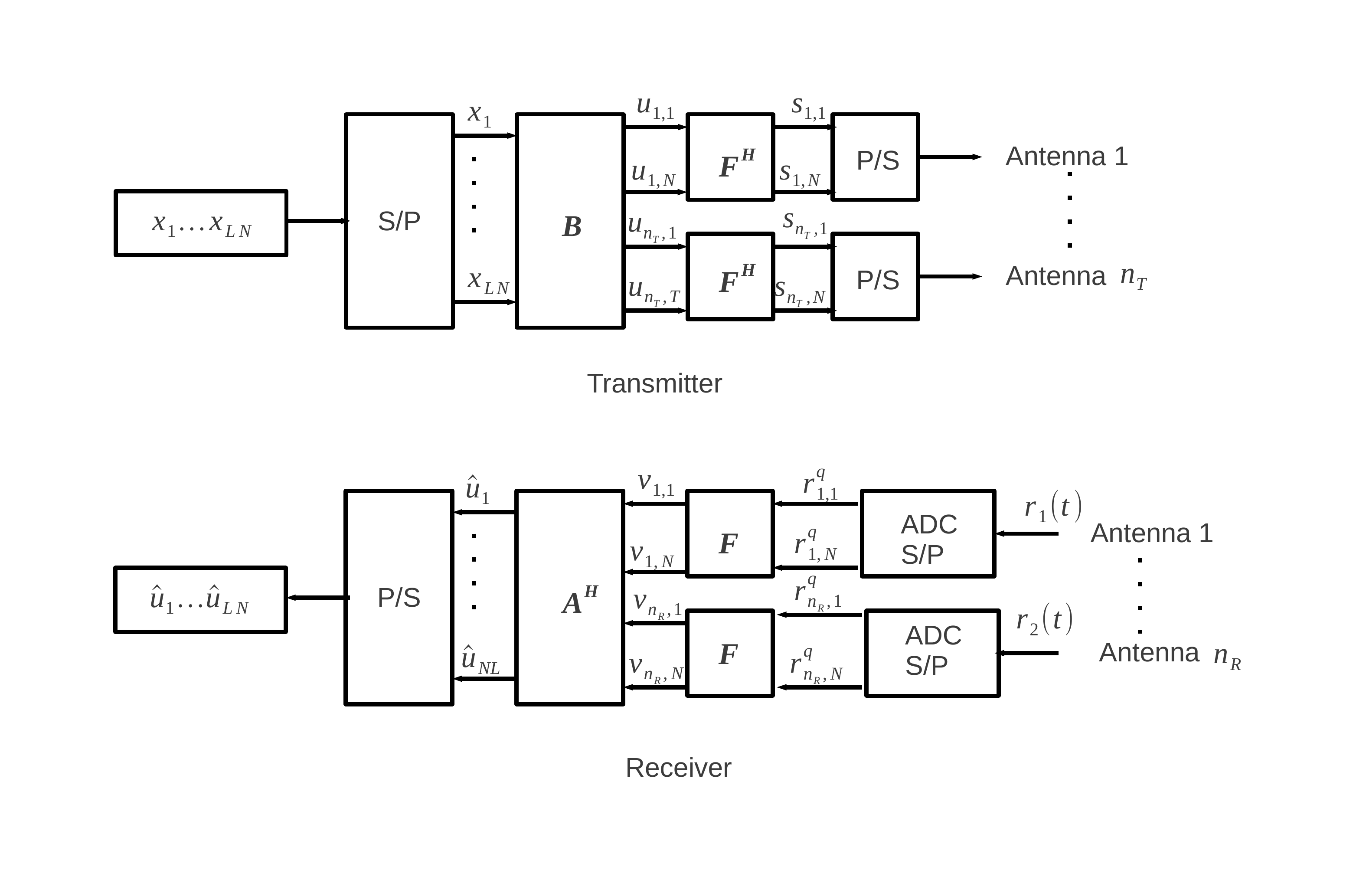}
\caption{System block diagram for MIMO OFDM with Tx-Rx beamforming.}
\end{figure}

We consider a communications system with $n_T$ transmit antennas and $n_R$ receiver antennas which gives rise to a MIMO channel. In case of flat faded channels, the MIMO channel is represented by a channel matrix, where any entry $(i,j)$ of the matrix is channel gain between antenna $i$ and antenna $j$. In case of MIMO frequency selective channel, a multicarrier scheme is often used and   each transmit antenna has an OFDM modulator and each receiver antenna has an OFDM demodulator (this can be assumed without any loss in capacity as showed in \cite{wang2000wireless,raleigh1998spatio}). A detailed explanation of single input single output (SISO) OFDM can be found in \cite{goldsmith2005wireless} and we omit several details details here. 

Let $\mathbf{u}_i\in\mathbb{C}^{N\times 1}$ be the frequency domain vector to be transmitted at antenna $i$. Define $\mathbf{u}:=\left[\mathbf{u}_1\ldots\mathbf{u}_{n_T}\right]^T$ and  $L:=\min(n_T,n_R)$. We assume a carrier--cooperative Tx--beamformer $\mathbf{B}\in \mathbb{C}^{n_TN\times LN}$ which allows for cooperation between different  subcarriers  while designing $\mathbf{B}$.  The  vector $\mathbf{u}$ is given by 
\begin{equation}
 \mathbf{u}=\mathbf{B}\mathbf{x}, \label{beamforming}
\end{equation}where $\mathbf{x}=\left[\mathbf{x}_1^T\ldots \mathbf{x}_L^T\right]^T\in\mathbb{C}^{LN}$  is the  data vector to be communicated. We assume w.l.o.g. $\text{E}[\mathbf{x}\mathbf{x}^\dagger]=\mathbf{I}$. Let $\mathbf{F}$ be a block diagonal matrix  of size $Nn_T\times Nn_T$ where each block is the  N-point discrete Fourier matrix $\mathbf{F_N}$. The time domain transmitted vector from any antenna $i$ is given by \begin{equation}\mathbf{s}_i=\mathbf{F}^\dagger_N\mathbf{u}_i \label{DFT}.\end{equation} Thus we can define a vector $\mathbf{s}:=\left[\mathbf{s}_1^T\ldots\mathbf{s}_{n_T}^T\right]^T=\mathbf{F}^\dagger\mathbf{u}$. The total power constraint at transmitter can be expressed as 

\begin{equation}
 \text{E}\left[||\mathbf{s}||^2\right]=Tr\left(\mathbf{B}^\dagger\mathbf{B}\right)\leq NL. \label{power_constraint}
\end{equation}

At the receiver, the analog samples are down converted and discretized. If the discretization is done at full precision, the received vector $\mathbf{r}_j\in\mathbb{C}^N$ (after removing the cyclic prefix) is given by
\begin{equation}
 \mathbf{r}_j=\mathbf{C}_j\mathbf{s}+\mathbf{w}_j, \label{received_antenna_unquant}
\end{equation}where $\mathbf{w}_j$ is additive zero mean Gaussian noise with covariance matrix $\xi^2\mathbf{I}$ and we define $\mathbf{C}_j:=[\mathbf{C}_{j,1}\ldots\mathbf{C}_{j,n_T}]$ and   $\mathbf{C}_{j,i}$ represents the time domain SISO channel between transmit antenna $i$ and  receive antenna $j$. The construction of the OFDM symbol forces the matrix $\mathbf{C}_{i,j}$ to be a circulant matrix. However, in practical systems, the discretization is done with finite precision. Let $A(\cdot)$ be the map which represents the analog-to-digital conversion. The ADC is defined by two parameters.

\noindent
 $Resolution$ $b$: If the resolution is $b$ bits, then the real and imaginary parts are each quantized to $2^b$ levels. 
 
 \noindent
$Range$: We assume that $A(\cdot)$ has a constant range of $(-1 ,+1 )$. If  the sampled signal exceeds this range, then it is clipped. In practice, an AGC block, with gain $G$ is used prior to the quantization to ensure that clipping occurs with low probability. In all our simulations, we use a uniform mid-point quantizer with range $(-1,1)$ and resolution $b$:
\begin{equation}
\begin{split}
A(x)&=\text{sign}\left(x\right)\left(\frac{1}{2^{b-1}}\lfloor 2^{b-1} |x|\rfloor+\frac{1}{2^b}\right),\quad |x|\leq 1.\\
&=\text{sign}\left(x\right)\left(1-\frac{1}{2^b}\right),\quad \text{otherwise}.
\end{split} \label{eq:quant}
\end{equation}where $\lfloor z \rfloor$ is the largest integer lesser than $z$. 
Then the received vector (after removing the cyclic prefix) at antenna $j$ is given by $
 \mathbf{r}^q_j=A\left(\mathbf{C}_j\mathbf{s}+\mathbf{w}_j\right), \label{received_antenna}
$where $A(\cdot)$ is applied elementwise. Defining $\mathbf{r}^q:=\left[\mathbf{r}^{q^T}_1\ldots\mathbf{r}^{q^T}_{n_R}\right]^T$,  and  $\mathbf{w}:=\left[\mathbf{w}^T_1\ldots\mathbf{w}^T_{n_R}\right]^T$ $\mathbf{C}=\left[\mathbf{C}_1\ldots\mathbf{C}_{n_R}\right]^T$, we can write
\begin{equation}
 \mathbf{r}^q=A\left(\mathbf{C}\mathbf{s}+\mathbf{w}\right) \label{quant_received}
\end{equation} 

\noindent
{\bf Modeling quantization noise:}
Due to the quantizer nonlinearity, analyzing an OFDM system with finite precision quantization becomes intractable. A simple heuristic is to model the quantization noise as additive and independent (see pseudo quantization noise model in Chapter 4 of \cite{q_noise}). It is shown in \cite{dardari} that the
PQN model is a valid model for quantization of OFDM signal only for a certain range of AGC. A description of the AGC calibration to ensure that the PQN model is valid is explained in \cite{tapan_tcom}. As per this description, if

\begin{equation}
 G^2=\frac{Nn_R\alpha}{\text{E}[||\mathbf{r}||^2]}, \label{AGC}
\end{equation}for a suitably chosen $\alpha$, where $\mathbf{r}=\left[\mathbf{r}_1^T\ldots \mathbf{r}_{n_RN}^T\right]^T$, the PQN model is a reasonable model for the quantization error.

\noindent
Using this model for quantization error, we can write 
\begin{equation}
 \mathbf{r}^q_j=\mathbf{C}_j\mathbf{s}+\mathbf{w}_j+\mathbf{q}_j .\label{received}
\end{equation}where $\mathbf{q}_j\in\mathbb{C}^{N\times 1}$ is the additive zero mean uniformly distributed quantization noise with covariance matrix $\frac{1}{G^2}\frac{2^{-2b}}{6}\mathbf{I}$. The receiver further transforms the received vector into the frequency domain by applying a $N$-point DFT, 

\begin{equation}
 \mathbf{v}_j=\mathbf{F}_N\mathbf{r}^q_j.
\end{equation} Combining \eqref{beamforming}, \eqref{DFT} and \eqref{received} along with the fact that $\mathbf{D}_{ji}:=\mathbf{F}_N\mathbf{C}_{ji}\mathbf{F}_N^\dagger$ is a diagonal matrix, we get

\begin{equation}
 \mathbf{v}_j=\mathbf{D}_j\mathbf{Bx}+\mathbf{F}_N\mathbf{w}_j+\mathbf{F}_N\mathbf{q}_j,
\end{equation}where $\mathbf{D}_j=\left[\mathbf{D}_{j1}\ldots \mathbf{D}_{jn_T}\right]$. Concatenating the frequency domain vectors from all $n_R$ antennas, we can write

\begin{equation}
 \mathbf{v}=\mathbf{DBx}+\mathbf{Fw}+\mathbf{Fq},
\end{equation}where $\mathbf{D}:=\left[\mathbf{D}_1\ldots \mathbf{D}_{n_R}\right]^T$, $\mathbf{v}:=\left[\mathbf{v}^T_1\ldots \mathbf{v}^T_{n_R}\right]^T$ and  $\mathbf{q}:=\left[\mathbf{q}^T_1\ldots \mathbf{q}^T_{n_R}\right]^T$. Let $\mathbf{A}$ be the Rx-beamformer. Again assuming cooperation among different carriers, the vector $\mathbf{v}$ is linearly transformed as 
\begin{equation}
 \hat{\mathbf{x}}=\mathbf{A}^\dagger\mathbf{v}=\mathbf{A}^\dagger\mathbf{DBx}+\mathbf{A}^\dagger\mathbf{Fw}+\mathbf{A}^\dagger\mathbf{Fq}. \label{final}
\end{equation} The statistic $\hat{\mathbf{x}}$ is used as a statistic to decode the data vector $\mathbf{x}$.

\section{Optimal Tx-beamforming for MIMO systems for a specified Rx-beamformer}

Often in downlink systems where the receivers have limited resources, it is beneficial to have a pre-specified linear receivers which depend only on the channel and not on the Tx-beamformer. For such systems systems, the goal is to design Tx-beamformers which optimizes a suitable metric. For most communication systems, the ultimate metric which we desire to optimize is the bit error rate. For the MIMO system \eqref{final} with $\hat{L}=\min\left(L,\text{rank}\left(D^\dagger D\right)\right)$ substreams, the average BER can be defined as 

\begin{equation}
 \text{BER}=\frac{1}{\hat{L}}\sum_{l=1}^{\hat{L}}\text{BER}_l
\end{equation}where $\text{BER}_l$ is the bit error rate for the $l$th substream. For a M-QAM constellation, a first order approximation of  $\text{BER}_l$  can be expressed as a function of the expected signal-to-interference noise ratio (SINR)  on the $l$th substream as

\begin{equation}
 \text{BER}_l\approx\frac{1}{\log_2M}\left(1-\frac{1}{\sqrt{M}}\right)Q\left(g_M\text{SINR}_l\right). \label{BER}
\end{equation}where $g_M=\frac{3}{M-1}$ and $Q(\cdot)$ is the tail probability of normal random distribution (\cite{palomar2003joint}). Using \eqref{final}, we can write

\begin{equation}
 \text{SINR}_l=\frac{|\mathbf{a}_l^\dagger\mathbf{Db}_l|^2}{\mathbf{a}^\dagger_l\left(\xi^2\mathbf{I}+\xi_q^2\mathbf{I}+\sum_{k\neq l}\mathbf{Db}_k\mathbf{b}_k^\dagger\mathbf{D}^\dagger\right)\mathbf{a}_l}
\end{equation}where $\xi_q^2=\frac{1}{G^2} \frac{2^{-2b}}{6}$ and $\mathbf{b}_l$ and $\mathbf{a}_l$ are the $l$th column vectors of $\mathbf{A}$ and $\mathbf{B}$, respectively.  Thus a BER minimizing criteria to design the Tx-beamformer can be written as

\begin{equation}\begin{split}
 \min_{\hat{\mathbf{B}}\in \mathbb{C}^{LN\times Nn_T}}\text{BER}, \quad \text{subject to}\\
 Tr\left(\hat{\mathbf{B}}^\dagger\hat{\mathbf{B}}\right)\leq NL. \label{BER_criteria}\end{split}
\end{equation} 

A standard method to solve such problems is to use the Lagrange multiplier method. However, this method does not provide any closed form solution to compute $\mathbf{B}$. Instead a set of matrix fixed point equations are obtained, the solution of which is difficult to compute over a large search space. In view of the space constraints, we do not write down this equations in this draft. For example, in a $2\times 2$ MIMO OFDM system with $128$ sub-carriers, we have $256\times 256$ variables to be optimized. For full precision systems,  several techniques have been used to get around the intractability of BER expressions e.g. using Chernoff bounds for the $Q(\cdot)$ function or maximizing the minimum of the SINR over all substreams. One other popularly used method is to minimize the mean square error (MSE) between $\mathbf{x}$ and $\hat{\mathbf{x}}$. We would like to point out that there exists a explicit analytical relationship between SINR and MSE only when optimal Weiner filters are used as  beamformers at 
the receiver. Therefore, maximizing the SINR is equivalent to minimizing the MSE only 
for jointly designing Tx-Rx beamformers. Although this techniques 
do not necessarily guarantee a closed form expression for the {\em optimal} Tx-beamformer, it often helps in designing simpler algorithms. As representative example, we explain one such  method  which minimizes the MSE.

\noindent
{\bf Minimizing the MSE criteria}: From \eqref{final}, the $\text{MSE}$ can be expressed  as 
\begin{equation}
 \text{MSE}=\text{E}\left[\|\mathbf{x}-\hat{\mathbf{x}}\|^2\right]=\|\mathbf{A}^\dagger\mathbf{DB}-\mathbf{I}\|^2_F+\left(\xi^2+\xi_q^2\right)\|\mathbf{A}\|_F^2,
\end{equation} where $\|\cdot\|_F$ denotes the Frobenius norm. For the moment we consider the full precision case i.e. $\xi_q^2=0$. Then, if the channel is perfectly known at the receiver and the transmitter,  the design criteria to find the optimal $\mathbf{B}$ is

\begin{equation}\begin{split}
 \min_{\hat{\mathbf{B}}\in\mathbb{C}^{Nn_T\times NL}}&\|\mathbf{A}^\dagger\mathbf{DB}-\mathbf{I}\|^2_F,\quad \text{subject to}\\ \label{MMSE}
Tr\left(\hat{\mathbf{B}}^\dagger\hat{\mathbf{B}}\right)&\leq NL.\end{split}
\end{equation}

\begin{lemma}
 Define $\bar{\mathbf{D}}:=\mathbf{A}^\dagger\mathbf{D}$. Let  $\mathbf{U}_{\bar{D}}\mathbf{\Delta}_{\bar{D}}\mathbf{V}_{\bar{D}}^\dagger$ and $\mathbf{U}_{B}\mathbf{\Delta}_{B}\mathbf{V}_{B}^\dagger$ be the singular value decompositions of $\bar{\mathbf{D}}$ and $\mathbf{B}$, respectively. Then optimality (with respect to \eqref{MMSE}) is achieved when $\mathbf{U}_B=\mathbf{V}_{\bar{D}}$ and $\mathbf{V}_B=\mathbf{U}_{\bar{D}}$. 
\end{lemma}This lemma is proved as a part of Theorem 1 in \cite{wang2009worst}. Consequently, this proves the optimality of eigen mode transmission and hence the optimality of the  diagonal structure of the complete channel described by the matrix $\mathbf{A}^\dagger\mathbf{DB}$. This reduces the complicated matrix optimization problem into a scalar power allocation problem, where the diagonal elements of the matrix $\mathbf{\Delta}_B$ gives the power allocated on the eigen modes. Using the above proposition, the optimization problem in \eqref{MMSE} simplifies to 

\begin{equation}\begin{split}
 \min_{\Delta_{B,1},\ldots,\Delta_{B,NL}}&\sum_{l=1}^{NL}\left(\Delta_{B,l}\Delta_{\bar{D},l}-1\right)^2,\quad \text{subject to}\\
 \sum_{l=1}^{NL}\Delta_{B,l}^2&\leq NL,\end{split} \label{MMSE}
\end{equation}where $\{\Delta_{B,l}\}$ and $\{\Delta_{\bar{D},l}\}$ are the diagonal elements of $\mathbf{\Delta}_{B}$ and $\mathbf{\Delta}_{\bar{D}}$, respectively.

\section{A simpler approach to design Tx-beamformers using the BER critieria}\label{section:sub-optimal}
As proved in Lemma 1, the diagonal structure is optimal while using minimum MSE as the  criterion for designing TX-beamformers for pre-specified Rx-beamformers. This simplification of  the problem makes it more amenable for obtaining closed form approximations or designing faster algorithms.  However, there does not exist a general {\em optimality of diagonalization} result for the BER or SINR criteria (diagonalization is optimal only when the Rx-beamformer is an optimal Weiner filter). Instead to utilize the useful properties of the diagonal structure, we suggest the following heuristic approach to design Tx-beamformers.

\begin{enumerate}
 \item For the specified linear Rx-beamformer $\mathbf{A}$, find the Tx-beamformer such that  $\mathbf{A}^\dagger\mathbf{DB}$ is diagonalized (as suggested in Lemma 1).
 \item Use this diagonal structure to obtain an expression of the average BER (which is the average of BER on each parallel sub-channel). 
 \item Compute the {\em optimal} eigen mode power allocation by minimizing the average BER.
\end{enumerate}As an example of the application of this approach, we consider a MIMO system where $\mathbf{A}=\mathbf{V}_D$ and $\mathbf{U}_D\mathbf{\Delta}_D\mathbf{V}_D$ is the singular value decomposition (SVD) of $\mathbf{D}$. According to   part 1) of the   method described above, we can use Lemma 1 to impose a diagonal structure on the complete MIMO system. This gives $\mathbf{B}=\sqrt{\mathbf{P}}\mathbf{U}_D$ where $\sqrt{\mathbf{P}}=\text{diag}\left(\sqrt{P_1}\ldots \sqrt{P_{NL}}\right)$ is the eigen power allocation to be determined. Such SVD based systems are often used in very high throughput systems (which is the main motivation of our work) which  try to maximize the multiplexing gain.
Without loss of generality, we assume all the singular values of $\mathbf{D}$ to be positive (If the matrix $\mathbf{D}$ has singular values to be zero, we remove that {\em parallel} channel from the system model).  Since $\mathbf{U}_D$ is unitary, the 
power constraint \eqref{power_constraint} is satisfied if $\text{Tr}\left(\mathbf{P}\right)\leq NL$. Under this structural assumptions, \eqref{final} can be written as

\begin{equation}
 \hat{\mathbf{x}}=\mathbf{P}\mathbf{\Delta_D}\mathbf{x}+\bar{\mathbf{w}}+\bar{\mathbf{q}}, \label{svd_final}
\end{equation}where $\bar{\mathbf{w}}=\mathbf{V}_D^\dagger\mathbf{Fw}$ and $\bar{\mathbf{q}}=\mathbf{V}_D^\dagger\mathbf{Fq}$. Since $\mathbf{V}$ and $\mathbf{F}$ are unitary, $\text{E}[\bar{\mathbf{w}}\bar{\mathbf{w}}^\dagger]=\xi^2\mathbf{I}$. Using the asymptotic normality results in \cite{shomorony_asymptotic_normality}, we can model
$\bar{\mathbf{q}}$ to be a a zero mean Gaussian vector with covariance matrix $\xi_q^2\mathbf{I}$ \footnote{Here we assume that any two  elements of vector $\bar{\mathbf{q}}$ are uncorrelated. This is not strictly true. However, this gives us simpler analytical expressions and at the same time gives accurate analytical predictions.}, where $\xi_q^2=\frac{1}{G^2} \frac{2^{-2b}}{6}$. Using the model \eqref{svd_final}, we have the following proposition

\begin{proposition}
 Under the preceding assumptions, the following statement holds true
 \begin{enumerate}
 \item The  uncoded BER for a $M-$QAM OFDM communication system with a $n_T\times n_R$ MIMO channel (parallelized into  $L=\min(n_T,n_R)$ independent channels as described in the preceding discussion) and eigen  power allocation  $\mathbf{P}$ is given by 
 \begin{equation}
  BER=\frac{4S-O(S^2)}{\log_2M}
\end{equation}where

\begin{equation*}\begin{split}
S&=\left(1-\frac{1}{\sqrt{M}}\right) \frac{1}{LN}\\ &\times \sum_{k=1}^{LN}Q\left(\sqrt{g_M\frac{P_k|\Delta_{D,k}|^2}{(c+1)\xi^2+\xi_q^2}}\right), \end{split}     
\end{equation*}

\begin{equation*}\begin{split}
g_M&=\frac{3}{M-1},
c=\frac{2^{-2b}}{6\alpha} \\ \text{and }&
 \xi_q^2= c\left(\frac{1}{LN}\sum_{j=0}^{LN}P_j|\Delta_{D,j}|^2\right).\end{split}
\end{equation*} and $\{\Delta_{D,k}\}$ are the singular values of $\mathbf{D}$.

\item An optimal eigenmode power allocation $\mathbf{P}^{(b)}$ (OEPA) which minimizes $S$ defined in Part 1)   is

\begin{equation}
P_k^{(b)}=\frac{\left(\xi^2+\xi_q^2\right)W\left(\frac{g_M|\Delta_{D,k}|^4}{\left((c+1)\xi^2+\xi_q^2\right)^2\left(\Omega^2+|\Delta_{D,k}|^2a^2\right)}\right)}{g_M|\Delta_{D,k}|^2},\label{optimal}
\end{equation}where $W(\cdot)$ is the principal value Lambert function \cite{lambert}, 
\[
a=c\sum_{j=1}^{LN}\frac{\sqrt{P^{(b)}_j|\Delta_{D,j}|^2}\exp\left(-\frac{g_MP_j^{b}|\xi_j|^2}{\xi^2+\xi_q^2}\right)}{\left(\xi^2+\xi_q^2\right)^\frac{3}{2}}
\] 
and
$\Omega$ is chosen to satisfy the power constraint \eqref{power_constraint}. The Lambert function $W(\cdot)$ is  defined as  inverse function of
$
f(w)=w\exp(w).
$

 \end{enumerate}

\end{proposition}  

{\bf Discussion}: The proof of part 1) and part 2) are on similar lines to the proof of part 1) and part 2) of Proposition 1 in \cite{tapan_tcom}. Computing OEPA using \eqref{optimal} is computationally expensive and we propose the following approximate OEPA (AOEPA). 

\begin{equation}
\tilde{P}_{k}^{(\infty)}= \frac{\frac{ W\left(\frac{g_M|\Delta_{D,k}|^4}{\xi^4 }\right)}{|\Delta_{D,k}|^2}}{\sum_{j=1}^{LN}\frac{ W\left(\frac{g_M|\Delta_{D,j}|^4}{\xi^4}\right)}{|\Delta_{D,j}|^2}},\quad \forall k, \label{eq:approximation_infinity}
\end{equation}

The motivation of the approximation  follows from the discussion in Section III.C of \cite{tapan_tcom}. 

{\bf Remark:} The ultimate goal is to minimize coded BER but for analytical tractability we have worked with uncoded BER. In the next section, we show using simulations that the proposed power allocation also improves coded BER performance.

\section{Simulation Results}

In this section, we present simulation results which highlights the improvement in the performance when using the Tx-Rx beamforming scheme presented in Section \ref{section:sub-optimal}. For carrying out the simulations,  the values of the parameters of the OFDM symbol and the MIMO channel model are summarized in Table \ref{table:parameter}.

\begin{table}[t!]
  \centering
  \caption{{Parameter values used in the simulation.}}
  \begin{tabular}{|c|c|c|}
  \hline
  Number of transmit antennas&$n_T$ &$2$\\ \hline
  Number of receive antennas& $n_R$ &$2$\\ \hline
  Name of parameter&Symbol &Value \\ \hline
  Number of subcarriers& $N$& $512$\\ \hline
  OFDM symbol duration& $T_s$  &$204.8$ ns \\ \hline
  Length of cyclic prefix & $L$ & 64 \\ \hline
  Cluster arrival rate&  $\Lambda$& $0.037$  $ns^{-1}$ \\ \hline
  Ray arrival rate& $\lambda$& $0.641$ $ns^{-1}$ \\ \hline
  Cluster decay rate& $\Gamma$&$21.1$ $ns$ \\ \hline
  Ray decay rate& $\gamma$ &$8.85$ $ns$ \\ \hline
  Cluster lognormal standard deviation& $\sigma_c$& $3.01$ dB \\ \hline
  Ray lognormal standard deviation& $\sigma_r$&$7.69$  dB \\ \hline
  Mean number of clusters & $L_c$&$3$\\ \hline
  Mean number of rays& $L_r$&$5$\\ \hline 
  \end{tabular} 
\label {table:parameter}
\end{table}

\begin{figure}[t!]
 \centering
 \includegraphics[trim=33mm 90mm 38mm 95mm,clip,width=3.4in]{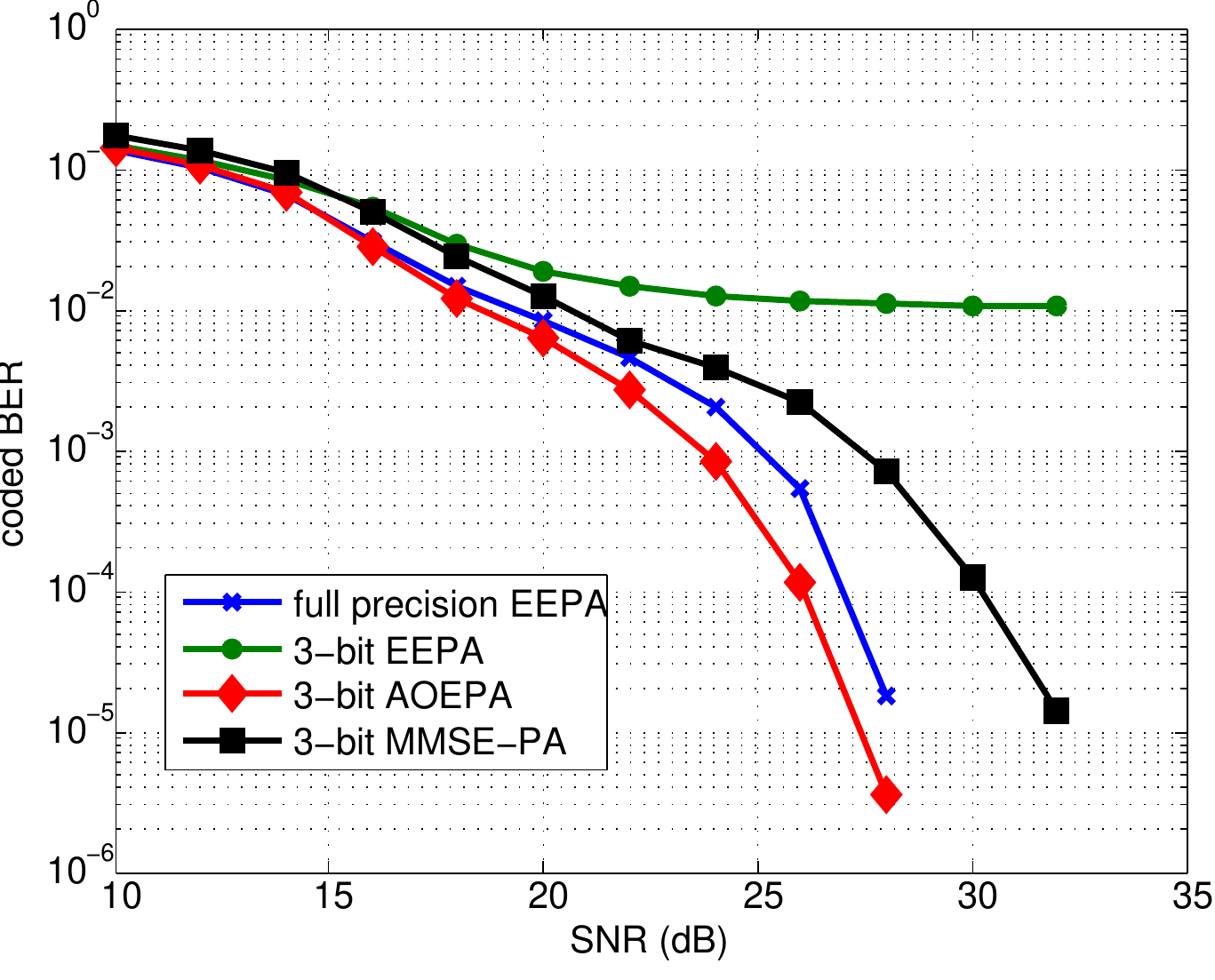}
 \caption{Coded BER performance for a rate $\frac{3}{4}-$rate LDPC coded $2\times 2$ high throughput MIMO OFDM system for EEPA, AOEPA and MMSE-PA.}
 \label{fig:performance}
\end{figure}

\begin{figure}[t!]
 \centering
 \includegraphics[trim=33mm 90mm 38mm 95mm,clip,width=3.4in]{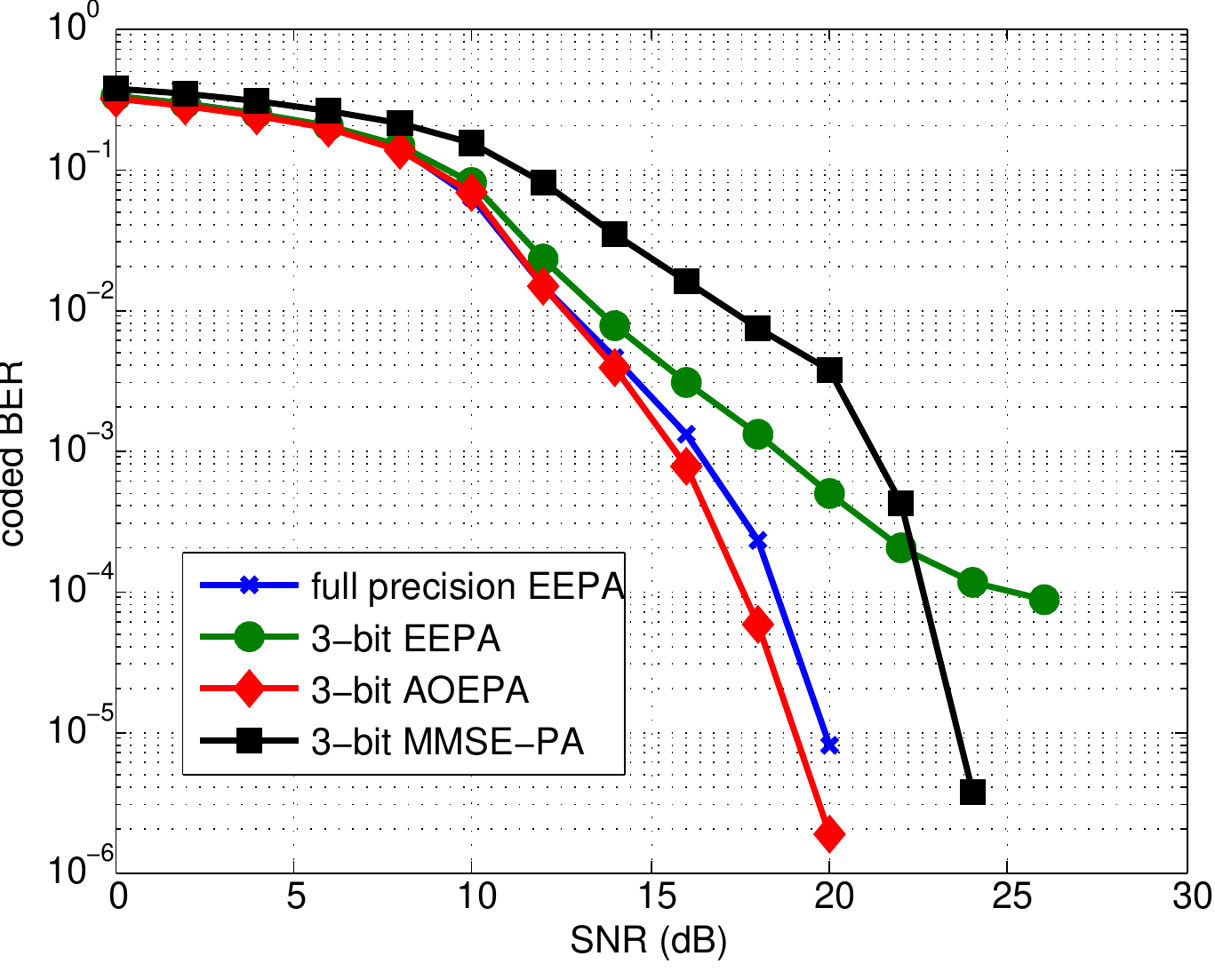}
 \caption{Coded BER performance for a rate $\frac{1}{2}-$rate LDPC coded $2\times 2$ high throughput MIMO OFDM system for EEPA, AOEPA and MMSE-PA.}
 \label{fig:performance1}
\end{figure}

 We consider a high throughput $2\times 2$ MIMO system communication system which is parallelized using  SVD described in previous section We highlight  four possible scenarios for such a system: 1) EEPA with full precision ADC, 2) EEPA with a 3-bit precision ADC, 3) MSE minimizing eigenmode power allocation (which we call MMSE-PA) obtained by solving \eqref{MMSE} with a 3-bit precision ADC, and 4) AOEPA given by \eqref{eq:approximation_infinity} with 3-bit precision ADC.  From Fig. \ref{fig:performance}, we see that for a $\frac{3}{4}$-rate low density parity check code (LDPC) coded  $2\times 2 $ MIMO OFDM system with 3-bit receiver, AOEPA achieves a BER of $10^{-4}$ at an SNR of $26$ dB compared to $27$ dB required with full precision with EEPA. On the other hand, a 3-bit system with EEPA requires has an error floor of $10^{-2}$. Similarly from Fig. \ref{fig:performance1}, for a $\frac{1}{2}$-rate $2\times 2 $ MIMO OFDM system, 3-bit AOEPA requires $1$ dB less power than EEPA full precision system. 
 
 The comparison of MSE minimizing eigen mode power allocation (MMSE-PA) with AOEPA 
provides a justification for our   method. Even though the MSE criteria has the {\em optimality of diagonalization} property while BER criteria is not guaranteed of such property,  optimal eigenmode power allocation which minimizes the MSE performs worse (it requires 30 dB to achieve BER of $10^{-4}$) compared to AOEPA. This justifies our approach to impose a diagonalizing structure on the MIMO system, computing the average BER (which is easier to calculate)  and thereafter computing the eigenmode power allocations which minimize the  BER.

\section{Conclusion}
In this paper, we investigate a Tx beamforming approach to improve performance of a  MIMO-OFDM system with a low
precision ADC at the receiver. We use Lemma 1 as a motivation and impose the structure emerging out of it to find a Tx-beamformer which minimizes the  BER criteria. The primary reason for imposing this structure is reduction of the dimensionality of the space we are optimizing over. Due to this structure,  the beamformer transmits  on the eigenmodes of the channel and power allocation on each eigenmode is the variable to be optimized.  For this structure, we compute the uncoded BER and find a eigenmode power allocation which minimizes the BER. We show that this eigenmode power allocation yields good performance compared to traditional systems with EEPA when   low  precision ADC are used at the receivers. In fact, our scheme achieves a performance which is comparable to that of full precision traditional systems.   As a part of future work, we would like to investigate the performance of our method for other commonly used receivers, {\em viz.} zero-forcing (ZF), minimum mean square error (MMSE)  and matched 
filters (MF). We would also like to gain further analytical 
insights into the {\em optimality of diagonalization} property for other design criteria.

\bibliographystyle{IEEEtran}
\bibliography{IEEEabrv,ref_journal}

\newcommand{\noop}[1]{}
\begin{thebibliography}{10}
\providecommand{\url}[1]{#1}
\csname url@samestyle\endcsname
\providecommand{\newblock}{\relax}
\providecommand{\bibinfo}[2]{#2}
\providecommand{\BIBentrySTDinterwordspacing}{\spaceskip=0pt\relax}
\providecommand{\BIBentryALTinterwordstretchfactor}{4}
\providecommand{\BIBentryALTinterwordspacing}{\spaceskip=\fontdimen2\font plus
\BIBentryALTinterwordstretchfactor\fontdimen3\font minus
  \fontdimen4\font\relax}
\providecommand{\BIBforeignlanguage}[2]{{%
\expandafter\ifx\csname l@#1\endcsname\relax
\typeout{** WARNING: IEEEtran.bst: No hyphenation pattern has been}%
\typeout{** loaded for the language `#1'. Using the pattern for}%
\typeout{** the default language instead.}%
\else
\language=\csname l@#1\endcsname
\fi
#2}}
\providecommand{\BIBdecl}{\relax}
\BIBdecl

\bibitem{ADCsurvey2008}
B.~Murmann, ``A/d converter trends: Power dissipation, scaling and digitally
  assisted architectures,'' in \emph{Custom Integrated Circuits Conference,
  2008. CICC 2008. IEEE}, 2008, pp. 105--112.

\bibitem{walden}
R.~Walden, ``Analog-to-digital converter survey and analysis,'' \emph{IEEE
  Journal on Selected Areas in Communications}, vol.~17, no.~4, pp. 539--550,
  Apr. 1999.

\bibitem{burmann}
B.~Murmann, ``A{DC} performance survey 1997-2010,'' [Online],
  http://www.stanford.edu/~murmann/adcsurvey.html.

\bibitem{tapan_vtc}
T.~Shah and O.~Dabeer, ``Subcarrier power allocation in {OFDM} with low
  precision {ADC} at receiver,'' in \emph{IEEE Vehicular Technology Conference
  (VTC Fall), 2012}, Sept., pp. 1--5.

\bibitem{tapan_tcom}
------, ``Optimal subcarrier power allocation for {OFDM} with low precision
  {ADC} at receiver,'' \emph{IEEE Transcations on Communications},
  \noop{3001}in press.

\bibitem{scaglione2002optimal}
A.~Scaglione, P.~Stoica, S.~Barbarossa, G.~B. Giannakis, and H.~Sampath,
  ``Optimal designs for space-time linear precoders and decoders,''
  \emph{Signal Processing, IEEE Transactions on}, vol.~50, no.~5, pp.
  1051--1064, 2002.

\bibitem{ding2003minimum}
Y.~Ding, T.~N. Davidson, Z.-Q. Luo, and K.~M. Wong, ``Minimum ber block
  precoders for zero-forcing equalization,'' \emph{Signal Processing, IEEE
  Transactions on}, vol.~51, no.~9, pp. 2410--2423, 2003.

\bibitem{palomar2003joint}
D.~P. Palomar, J.~M. Cioffi, and M.~A. Lagunas, ``Joint tx-rx beamforming
  design for multicarrier mimo channels: A unified framework for convex
  optimization,'' \emph{Signal Processing, IEEE Transactions on}, vol.~51,
  no.~9, pp. 2381--2401, 2003.

\bibitem{wang2000wireless}
Z.~Wang and G.~B. Giannakis, ``Wireless multicarrier communications,''
  \emph{Signal Processing Magazine, IEEE}, vol.~17, no.~3, pp. 29--48, 2000.

\bibitem{wpan}
\emph{IEEE Std 802.15.3c-2009 (Amendment to IEEE Std 802.15.3-2003)}, pp.
  c1--187, Oct. 2009.

\bibitem{ch_par}
S.~Yong, ``Channel model sub-committee final report,''
  https://mentor.ieee.org/802.15/file/07/15-07-0584-01-003c-tg3c-channel-model%
ing-sub-committee-final-report.doc, March 2007.

\bibitem{ch_par1}
------, \emph{60 GHz Technology for Gbps WLAN and WPAN}.\hskip 1em plus 0.5em
  minus 0.4em\relax John Wiley \& Sons, Ltd, 2010, pp. 17--61.

\bibitem{raleigh1998spatio}
G.~G. Raleigh and J.~M. Cioffi, ``Spatio-temporal coding for wireless
  communication,'' \emph{Communications, IEEE Transactions on}, vol.~46, no.~3,
  pp. 357--366, 1998.

\bibitem{goldsmith2005wireless}
A.~Goldsmith, \emph{Wireless communications}.\hskip 1em plus 0.5em minus
  0.4em\relax Cambridge university press, 2005.

\bibitem{q_noise}
B.~Widrow and I.~Koll\'ar, \emph{Quantization Noise: Roundoff Error in Digital
  Computation, Signal Processing, Control, and Communications}.\hskip 1em plus
  0.5em minus 0.4em\relax Cambridge, UK: Cambridge University Press, 2008.

\bibitem{dardari}
D.~Dardari, ``Joint clip and quantization effects characterization in {OFDM}
  receivers,'' \emph{IEEE Transactions on Circuits and Systems I: Regular
  Papers}, vol.~53, no.~8, pp. 1741--1748, Aug. 2006.

\bibitem{wang2009worst}
J.~Wang and D.~P. Palomar, ``Worst-case robust mimo transmission with imperfect
  channel knowledge,'' \emph{Signal Processing, IEEE Transactions on}, vol.~57,
  no.~8, pp. 3086--3100, 2009.

\bibitem{shomorony_asymptotic_normality}
I.~Shomorony and A.~S. Avestimehr, ``Worst-case additive noise in wireless
  networks,'' \emph{Computing Research Repository}, vol. abs/1202.2687, 2012.

\bibitem{lambert}
A.~Hoorfar and M.~Hassani, ``Inequalities of {L}ambert {W} function and
  hyperpower function,'' \emph{Journal of inequalities in pure and applied
  mathematics}, vol.~9, 2008.

\end{thebibliography}
%

\end{document}